\documentclass[journal]{IEEEtran}
\ifCLASSINFOpdf
\else
\fi
\usepackage{amsmath}
\usepackage{graphicx}
\usepackage{mdwtab}
\usepackage{booktabs}
\usepackage{multirow}
\usepackage{multicol}
\usepackage{xcolor}
\usepackage{amssymb}
\usepackage{hyperref}
\usepackage[ruled,vlined]{algorithm2e}
\usepackage[caption=false,font=normalsize,labelfont=sf,textfont=sf]{subfig}
\usepackage{textcomp}
\usepackage{stfloats}
\usepackage{adjustbox}

\newcommand\Y{\mathbf Y}
\newcommand\E{\mathbf E}
\newcommand\A{\mathbf A}

\newcommand\D{\mathbf D}

\newcommand\B{\mathbf B}

\newcommand\bS{\mathbf S}
\newcommand\bL{\mathbf L}

\newcommand\I{\mathbf I}
\newcommand\Q{\mathbf Q}

\newcommand\1{\bf 1}

\newcommand{\Lagr}{\mathcal{L}}

\begin{document}
%
\title{MiSiSUn: Minimum Simplex Semisupervised Unmixing} 
%
%
%

\author{Behnood~Rasti,~\IEEEmembership{Senior~Member,~IEEE,}
Bikram~Koirala,~\IEEEmembership{Member,~IEEE,}
Paul~Scheunders, ~\IEEEmembership{Senior~Member,~IEEE}
\thanks{Behnood Rasti (corresponding author) is with  the Faculty of Electrical Engineering and Computer Science, Technische Universität Berlin, 10623 Berlin, Germany, and also with the Berlin Institute for the Foundations of Learning and Data (BIFOLD), 10623 Berlin, Germany; behnood.rasti@tu-berlin.de, behnood.rasti@gmail.com}
\thanks{Bikram Koirala and Paul Scheunders are with	the Imec-VisionLab, Department of Physics, University of Antwerp, Belgium (email: bikram.koirala@uantwerpen.be; paul.scheunders@uantwerpen.be).}
\thanks{Manuscript received  2026; revised 2026.}}

%
%

\markboth{Journal of \LaTeX\ Class Files,~Vol.~?, No.~?, ?~2026}%
{Shell \MakeLowercase{\textit{et al.}}: Bare Demo of IEEEtran.cls for Journals}
%



\maketitle

\begin{abstract}
This paper proposes a semisupervised geometric unmixing approach called minimum simplex semisupervised unmixing (MiSiSUn). The geometry of the data was incorporated for the first time into library-based unmixing using a simplex-volume-flavored penalty based on an archetypal analysis-type linear model. The experimental results were performed on two simulated datasets considering different levels of mixing ratios and spatial instruction at varying input noise. MiSiSUn considerably outperforms state-of-the-art semisupervised unmixing methods. The improvements vary from 1 dB to over 3 dB in different scenarios. The proposed method was also applied to  a real dataset where visual interpretation is close to the geological map. MiSiSUn was implemented using PyTorch, which is open-source and available at \href{https://github.com/BehnoodRasti/MiSiSUn}{https://github.com/BehnoodRasti/MiSiSUn}. Moreover, we provide a dedicated Python package for Semisupervised Unmixing, which is open-source and includes all the methods used in the experiments for the sake of reproducibility.
\end{abstract}

\begin{IEEEkeywords}
Library-based unmixing, sparse unmixing, hyperspectral, sparsity, semi-supervised, blind, unmixing, PyTorch, GPU, alternating direction method of multipliers, nonconvex, optimization
\end{IEEEkeywords}

%
\IEEEpeerreviewmaketitle

\section{Introduction}

\IEEEPARstart{S}{pectral} unmixing is a crucial technique in hyperspectral remote sensing, vital for applications such as mineral exploration, agriculture, environmental monitoring, urban planning, planetary surface analysis, pollution monitoring, medical imaging, and water quality assessment. Hyperspectral sensors capture contiguous spectra, enabling the separation and identification of different materials within an image. Unmixing algorithms decompose mixed spectral data into its constituent parts using endmembers, which are unique spectral signatures of materials. However, the process is challenging due to low spatial resolution, multiple scattering, and intimate mixing, which result in complex spectra within each pixel \cite{HySUPP}. 

In hyperspectral remote sensing, a mixing model represents the observed spectral pixel as a combination of endmembers and their fractional abundances. Unmixing involves estimating these fractional abundances by either identifying the endmembers directly from the data or using a pre-existing library.  It may also require determining the number of endmembers present. The mixing model can be linear or nonlinear, depending on the interactions between incident light and the materials in the scene \cite{HySUPP}. This paper discusses library-based linear unmixing models found suitable for Earth observation \cite{unmixing-review, HySUPP}.

There are three main strategies to tackle the linear unmixing problem. 1) Identifying endmembers and estimating abundances sequentially, i.e., first, extracting/estimating endmembers (often by using a geometric approach \cite{MParente_2010_review}) and then estimating the abundances using an abundance estimation technique such as fully constrained least squares unmixing (FCLSU) \cite{FCLSU}. This was referred to as an unmixing chain \cite{unmixing-review} due to the sequential nature of the process or as supervised unmixing \cite{HySUPP} due to the assumption that endmembers are known while estimating the abundances. 
Geometric approaches are often successful when either pure pixels exist for materials within the data or sufficient spectra are located on the edges of the data simplex. As a result, endmembers can be successfully extracted or estimated, which plays the main role in successful abundance estimation.
2) Estimating endmembers and abundances simultaneously, referred to as blind (unsupervised) unmixing. Statistical approaches or shallow/deep neural networks were often used for blind unmixing, when no pure pixels are available in the data. In highly mixed scenarios, however, they often fail due to the large sets of solutions that fit the data \cite{HySUPP, unmixing-review}. 
3) Library-based (semi-supervised) sparse unmixing approaches estimate abundances by relying on an endmember spectral library. To be successful in highly mixed scenarios, the library should be improved by removing similar spectra (e.g.,  scaled versions) of the true endmembers of the dataset.
In this paper, we propose an approach within the latter category that does not require improving the spectral library. Therefore, the next section discusses different library-based approaches. \textcolor{black}{More importantly, we discuss a family of library-based algorithms that neither entirely rely on an endmember library nor blindly estimate the endmembers, but incorporate an endmember library to estimate endmembers and abundances simultaneously. }


J. M. Bioucas-Dias and M. A. T. Figueiredo reformulated library-based unmixing by introducing sparse regression. They introduced Sparse Unmixing by Variable Splitting and Augmented Lagrangian (SUnSAL) and its variant, Constrained SUnSAL (C-SUnSAL) \cite{SUnSAL}. Both SUnSAL and C-SUnSAL employ the $\ell_1$ penalty to promote sparsity in abundance estimation. SUnSAL minimizes the $\ell_1$ norm additional to the $\ell_2$ data fidelity term, while C-SUnSAL uses the $\ell_2$ data fidelity term as a constraint to minimize the $\ell_1$ norm. To incorporate spatial information, in \cite{SUnSAL-TV}, total Variation was added to the $\ell_1$ norm (SUnSAL-TV). In \cite{CoSUn}, collaborative sparse unmixing (CoSUn) was proposed, which exploits a group constraint by applying the sum of $\ell_2$ norms to the abundances. Later on, the selection of a suitable regularizer for sparse unmixing became a big challenge. Spatial-spectral weighted $\ell_1$ norms were widely explored for sparse unmixing \cite{SZhang2016, SZhang2018, DRSU_TV,SU_LR, MRS2WSU}. Local collaborative sparse unmixing (LCSU) \cite{SZhang2016} and spectral-spatial weighted sparse unmixing (S$^2$WSU) \cite{SZhang2018} were proposed to incorporate spatial information and better promote sparsity on abundances. In \cite{DRSU_TV}, non-isotropic TV was integrated with a spatial-spectral weighted $\ell_1$ norm. In \cite{SU_LR}, a weighted nuclear norm was combined with the weighted $\ell_1$ norm to leverage the low-rank properties of the abundances. Some methods were proposed to exploit endmember bundles and structured dictionaries to address the spectral variability in the spatial domain \cite{GSIMN, MBSUn}.  In \cite{GSIMN},  sparse unmixing was proposed using a group sparsity norm relying on a dictionary of endmember bundles generated from the data points. Tensor decomposition was also used with a sparse unmixing formulation using mode-n tensor multiplication \cite{TD_SU1, TD_SU2}.

To remove spectral variation and capture spatial information, segmentation-based sparse unmixing approaches were suggested.  In \cite{RABorsoi2019}, a fast Multiscale Sparse Unmixing Algorithm (MUA) was proposed, which compares three segmentation techniques: binary partition tree (BPT), simple linear iterative clustering (SLIC), and K-means clustering to group pixels. SUnSAL was applied to the pixel average of every segment to estimate coarse fractional abundances, ultimately used as a prior for the final sparse regression. This concept is also known in the literature as "super-pixel", and several methods have been developed under this terminology. In \cite{TanerInce2020}, SLIC was used for pixel grouping, and pixel-based sparse unmixing was performed using a superpixel-based graph Laplacian regularization.

Deep Neural networks were also developed for sparse unmixing. In \cite{SUnCNN}, a convolutional neural network (SUnCNN) was proposed for sparse unmixing. The sparse regression optimization was transformed into an optimization task involving the parameters of a deep encoder-decoder network, while the ASC could be enforced using a softmax layer.  In \cite{SMALU}, an asymmetric encoder-decoder network is used with a sparse variation of softmax to avoid the full support of softmax while enforcing ASC. Generally speaking, selecting suitable hyperparameters for such deep networks is often a challenging endeavor.

 Algorithm unrolling-based methods were also explored for sparse unmixing \cite{ISTA_Unrol_SUn, SU_Unroll}. In \cite{ISTA_Unrol}, the Iterative Soft-Thresholding Algorithm (ISTA) \cite{ISTA} was unrolled to tackle the sparse unmixing problem.  In \cite{SUNSAL_Unrol}, a network was proposed based on the unrolling of the SUnSAL algorithm.  An intermediate convolutional layer was applied to the abundance representation to enhance the integration of spatial information.

 The main disadvantage of conventional sparse unmixing methods mentioned above is that they cannot incorporate geometric information on the fixed large spectral library, and are therefore not efficient in highly mixed scenarios. On the other hand, when the number of pixels of the HSI increases, they become inefficient.
 
 In this paper, we propose a semisupervised unmixing approach called Minimum Simplex Semisupervised Unmixing (MiSiSUn), which addresses the above-mentioned drawbacks by using a library-based archetypal analysis model. In archetypal analysis, a measured spectrum is modeled as a mixture of archetypes (extremes of the dataset). When no pure pixels are present in the dataset, archetypes are themselves mixtures of endmembers. This challenge is addressed by exploiting the geometric information of the linear simplex by enforcing a minimum volume-flavored penalty to the constrained least squares optimization. Furthermore, we propose a GPU-accelerated implementation using PyTorch to improve efficiency.  
 

The remaining sections of this paper are organized as follows. Section II discusses some related works. Section III describes the proposed MiSiSUn. Section IV presents the experimental results. Section V concludes this paper.






\section{Related Works}
\subsection{Linear Mixing Model and Endmember Variability}


The conventional (low-rank) linear mixing model for an observed hyperspectral pixel ${\bf y} \in  \mathbb{R}^{p}$ (i.e., the sensor has $p$ bands), is given by:
 \begin{equation}\label{eq: M0}
{\bf y} = {\bf E}{\bf a} + {\bf n}, ~~{\rm s.t.~~}\sum_{i=1}^ra_i=1, a_i\geq 0, i=1,2,..,r,
\end{equation} 
where ${\bf E}\in \mathbb{R}^{p\times r}$ and ${\bf a}^T$ $\in \mathbb{R}^{r}$ denote the  $r$ endmembers and their fractional abundances, respectively. ${\bf n}$ is a $p$-dimensional random vector denoting  additive random Gaussian noise. For $n$ pixels, we have:
 \begin{equation}\label{eq: LR}
{\bf Y} = {\bf E}{\bf A} + {\bf N}, ~~{\rm s.t.~~}{\bf A}\geq 0,{\bf 1}_{r}^{T}{\bf A}={\bf 1}_{n}^{T},
\end{equation} 
where   {\bf Y}$ \in  \mathbb{R}^{p\times n}$ is the observed HSI,  containing $p$ bands and  $n$ pixels, ${\bf A}$ $\in \mathbb{R}^{r\times n}$  contains the fractional abundances, ${\bf N}$  $\in \mathbb{R}^{p\times n}$ is noise. ${\bf 1}_n$ indicates an $n$-component column vector of ones. LMM is often used for supervised and blind unmixing approaches. It has a geometric interpretation, i.e., the noise-free data points are located within a ($r-1$) simplex, and abundances must be within ($r-1$) probability simplex ($\Delta_r$) \cite{HySUPP}. This geometric interpretation led to a large group of unmixing approaches often known as geometric approaches \cite{unmixing-review}. However, these approaches cannot take into account the spectral/endmember variabilities that occur during hyperspectral acquisitions. Such variabilities are often caused by noise, atmospheric effects, topography (e.g., occlusion of light), and intrinsic variability of materials (\cite{Spec_var}). To take into account the endmember variability, which can partially address the spectral variability, multiple endmember spectral and
mixture analysis (MESMA) \cite{MESMA} was proposed. MESMA allows every pixel to have different endmembers, relying on a structured library, $\D=[\D_1,\D_2, ..., \D_r]$, containing the endmember bundles ${\bf D}_i$, one for each material $i$. Assuming the pixel-wise LMM, MESMA solves: 
\begin{align}\nonumber
      &\hat{\bf a}_i,\hat{\bf E}_i=\arg\min_{{\bf a}_i, {\bf E}_i} \frac{1}{2} || {\bf y}_i-{\bf E}_i{\bf a}_i||_{2}^{2}
~~~{\rm s.t.}~~~\\ &{\bf E}_i\in{\bf D}, {\bf a}_i\geq 0,{\bf 1}_{r}^{T}{\bf a}_i=1.
\end{align}
To solve this combinatorial problem, MESMA seeks all combinations of endmembers within endmember bundles and selects the combination having the lowest reconstruction error. Considering such a solution for all the pixels, MESMA is a highly computationally demanding algorithm. We should note that this search can be performed in parallel for all the pixels. Nevertheless, MESMA is computationally costly, and therefore, several algorithms have been proposed to reduce the computational burden \cite{MESMA,End_Rob, Comp_MESMA, MELSUM}.

\subsection{Sparse and Redundant Model and Sparse Unmixing}
The emergence of the sparse and redundant models (in the early 2000s) affected the signal and image processing field. Consequently, those models had a significant impact on hyperspectral processing, including unmixing. The sparse and redundant linear model (SRM) is given by:
 \begin{align}\label{eq: SR}\nonumber
&{\bf Y} = {\bf D}{\bf X} + {\bf N}, \\&
~~~{\rm s.t.}~~~{\bf X}\geq 0,{\bf 1}_{m}^{T}{\bf X}={\bf 1}_{n}^{T},  
\end{align} 
where $ {\bf D}  \in \mathbb{R}^{p\times m} $  ($p\ll m$) denotes the spectral library containing $m$ endmembers and $ {\bf X}\in \mathbb{R}^{m\times n}$ is the unknown  fractional abundances to estimate. Please note that ${\bf D}$ serves as an overcomplete dictionary, and as such, it should be carefully crafted for unmixing. A well-structured dictionary comprises endmembers representing the materials present in the scene and can efficiently reduce the redundancy in ${\bf X}$. Using a well-designed dictionary, each pixel in the scene is represented by a combination of only a few dictionary atoms, resulting in 
${\bf X}$ being a sparse matrix. If a particular endmember material is absent from the observed spectra, the corresponding row in ${\bf X}$ will be entirely zero. This sparsity is common because abundance values in this model are typically sparse. 
J. M. Bioucas-Dias and M. A. T. Figueiredo solved the sparse and redundant model  (\ref{eq: SR})  by \cite{SUnSAL}:
\begin{align}\label{eq: SUnSAL}\nonumber
  &\hat{\bf X}=\arg\min_{{\bf X}} \frac{1}{2} || {\bf Y}-{\bf DX}||_{F}^{2}+\lambda ||{\bf X}||_1\\&
~~~{\rm s.t.}~~~{\bf X}\geq 0,{\bf 1}_{m}^{T}{\bf X}={\bf 1}_{n}^{T}.
\end{align}
This framework is frequently used in sparse unmixing, where fractional abundances 
${\bf X}$ are estimated by applying sparsity-enforcing penalties or constraints within a sparse regression formulation \cite{SUn}. This solved the lack of efficiency in MESMA and opened an active research field called sparse unmixing \cite{SUnSAL_Leg}. However, despite their efficacy, many sparse unmixing approaches often lack geometric and physical interpretations. There is debate as to whether ASC should be applied in enforcing sparsity. Most sparse unmixing methods that exploit the $\ell_1$ norm do not use ASC since those constraints conflict. For instance, SUNSAL with ASC turns into FCLSU since $\ell_1$ does not play any role after enforcing ASC. However, some sparse unmixing methods that exploit other sparsifying norms such as the $\ell_q$ norm also enforce ASC \cite{SU_lq, GSIMN}. In \cite{SUn}, the difference between enforcing ASC and ignoring it was analyzed for classical sparse regression-based techniques, where in many cases omitting ASC outperforms enforcing it. The comparisons were based on SRE and simulated datasets with different noise scenarios. We should note that ignoring ASC breaks physical constraints on pixels for the mixture model.

\subsection{Archetypal Analysis Model and Semisupervised Unmixing}
Archetypal Analysis Models (AAM) \cite{RAA}  were proposed for blind unmixing \cite{cutler1994archetypal, EDAA}. The AAM is given by: 
\begin{align}\label{eq: AA}\nonumber
 &{\bf Y}={\bf YBA}+{\bf N} \\&
{\rm s.t.}~~{\bf B}\geq 0,{\bf 1}_{n}^{T}{\bf B}={\bf 1}_{r}^{T}, {\bf A}\geq 0,{\bf 1}_{r}^{T}{\bf A}={\bf 1}_{n}^{T},
\end{align}
where ${\bf B} \in \mathbb{R}^{n\times r}$ and the columns of  ${\bf B}$ belong to the simplex $\Delta_n$. In other words, the endmembers or archetypes are convex combinations of the data points. This is a strict constraint, meaning that the endmembers are close to data points, which may not be the case for unmixing applications, particularly for highly mixed scenarios. 

Inspired by the AAM (\ref{eq: AA}), in \cite{SUnAA}, Rasti et. al. proposed a library-based AAM for semisupervised unmixing:
 \begin{equation}\label{eq: M2}
{\bf Y} = {\bf D}{\bf B}{\bf A} + {\bf N}. 
\end{equation} 
 where ${\bf E}={\bf DB}$ is a convex combination of library endmembers, i.e., the columns of ${\bf D}$, which makes the model suitable for highly mixed scenarios. 
Even a carefully pruned and well-selected spectral library cannot represent all the endmembers of materials in a real-world dataset. Factors such as noise, atmospheric effects, illumination variations, and the intrinsic variability of materials can alter the endmembers and introduce scaling factors when comparing the scene to the library endmembers. These discrepancies are known as library mismatches. This issue was partially addressed in \cite{SUn} by applying a bandwise scaling factor to the observed data. In \cite{SUnAA}, SUnAA was proposed, which solves a nonconvex optimization to  simultaneously estimate ${\bf B}$ and ${\bf A}$:
\begin{align}\label{eq: M33}\nonumber
  &(\hat{\bf B},\hat{\bf A})=\arg\min_{{\bf B,A}} \frac{1}{2} || {\bf Y}-{\bf DBA}||_{F}^{2} \\&
{\rm s.t.}~{\bf B}\geq 0,{\bf 1}_{m}^{T}{\bf B}={\bf 1}_{r}^{T},  {\rm and}~{\bf A}\geq 0,{\bf 1}_{r}^{T}{\bf A}={\bf 1}_{n}^{T}.
\end{align}
A parameter-free solution was derived to solve (\ref{eq: M33}) using active set methods. However, SUnAA runs on a CPU and is highly time-consuming. Therefore, in \cite{FaSUn}, an ADMM-based solution called FaSUn was given for the proposed minimization problem  (\ref{eq: M33}), which benefits from a GPU-accelerated implementation. In \cite{FUnmix}, a sparse  unmixing using soft-shrinkage (SUnS), which enforces $\ell_1$ on $\B$ was proposed, given by 
\begin{align}\label{eq: SUnS1}\nonumber
  &(\hat{\bf B},\hat{\bf A})=\arg\min_{{\bf B,A}} \frac{1}{2} || {\bf Y}-{\bf DBA}||_{F}^{2} +\lambda ||{\bf B}||_1 \\&
{\rm s.t.}~{\bf A}\geq 0,{\bf 1}_{r}^{T}{\bf A}={\bf 1}_{n}^{T}, 0\leq{\bf B}\leq 1.
\end{align}
The comparative experiments in \cite{FUnmix} revealed that FaSUn outperforms SUnS, showing that the convexity constraint applied to $\B$ is more efficient. 

Here, we summarize the pros and cons of AAM (\ref{eq: M2}) compared to LMM (\ref{eq: LR}) and SRM (\ref{eq: SR}):
 \begin{itemize}
     \item Both LMM and SRM assume that the (unknown) endmembers are present in the library, which is unrealistic. Often, the true endmembers for the observed dataset are scaled versions of the library endmembers due to noise and spectral variations introduced during data acquisition. On the other hand, AAM assumes that endmembers are convex combinations of the library endmembers. 
     \item AAM has a geometric interpretation compared to SRM. Therefore, it allows the incorporation of geometric information when developing algorithms.
     \item AAM has a physical interpretation and allows ASC. Note that sparse unmixing approaches often omit this constraint. \item AAM is non-convex, and the number of endmembers should be known, which makes it more challenging to solve compared to SRM.  
     \item AAM is a low-rank model that decreases the computational time.
     \end{itemize}

\section{MiSiSUn: Minimum Simplex Semisupervised Unmixing}
\label{subsec: FaSUn}
When pure pixels are unavailable in the datasets, AAM  combines library spectra to generate virtual endmembers for the dataset. The generated virtual endmembers may not necessarily correspond to the real endmembers of the dataset. To reliably generate virtual endmembers, it is necessary to constrain the volume of the data simplex. In this work, we design a method that integrates the properties of AAM and the linear simplex. In particular, we add the so-called center penalty \cite{CoNMF, NMF_QMV, MiSiCNet}, which constrains the volume of the simplex in the least-squares term used in AAM.
The resulting nonconvex optimization to  simultaneously estimate ${\bf B}$ and ${\bf A}$ is:
\begin{align}\label{eq: M3}\nonumber
  &(\hat{\bf B},\hat{\bf A})=\arg\min_{{\bf B,A}} \frac{1}{2} || {\bf Y}-{\bf DBA}||_{F}^{2} +\lambda\left\| {\bf DB} - {\bf m}{\bf 1}_{r}^{T} \right\|_F^2\\&
{\rm s.t.}~~{\bf B}\geq 0,{\bf 1}_{m}^{T}{\bf B}={\bf 1}_{r}^{T},  {\rm and }~~{\bf A}\geq 0,{\bf 1}_{r}^{T}{\bf A}={\bf 1}_{n}^{T}.
\end{align}
where, ${\bf m}$ contains the mean values of the spectral pixels, i.e., ${\bf m}=\frac{1}{n}{\bf Y}{\bf 1}_n$.  This term pulls the endmembers toward the center of mass. 

Here, we propose an ADMM-based solution for the minimization problem  (\ref{eq: M3}), which benefits from a GPU-accelerated implementation. 
First, we should note that the minimization problem (\ref{eq: M3}) is non-convex; however, it can be solved in two steps using a cyclic descent scheme; ${\bf A}$-step: when ${\bf B}$ is fixed and ${\bf B}$-step: ${\bf A}$ is fixed. In every step, we are dealing with a convex optimization, and therefore, the solution of every step successively decreases the loss function, leading to a minimum. The convergence of the final solution is guaranteed upon the convergence of every step.

\subsection{${\bf A}$-step:} In the first step, we assume that ${\bf B}$ is fixed and therefore ${\bf E}={\bf DB}$ is fixed. Then, problem (\ref{eq: M3}) turns into a least squares problem with a simplex constraint given by:
\begin{align}\label{eq: FCLSU}\nonumber
  &\hat{\bf A}=\arg\min_{{\bf A}} \frac{1}{2} || {\bf Y}-{\bf EA}||_{F}^{2} \\&
{\rm s.t.}{\bf A}\geq 0,{\bf 1}_{r}^{T}{\bf A}={\bf 1}_{n}^{T}.
\end{align}
Problem (\ref{eq: FCLSU}) can be solved using various methods such as convex optimization, least squares, or quadratic programming solvers. However, in unmixing and Earth observation applications, we often deal with large datasets, making these general-purpose solvers inefficient. To address this, we propose an ADMM-based solution.

To solve problem  (\ref{eq: FCLSU}), we start by splitting ${\bf A}$: 
    \begin{align}\label{eq: FCLSU2}\nonumber
 & \hat{\bf A},\hat \bS=\arg\min_{{\bf A},{\bS}} \frac{1}{2} || {\bf Y}-{\bf E}{\A}||_{F}^{2}\\ &~~ {\rm s.t.}~~ \A=\bS,~~
{\bf S}\geq 0,{\bf 1}_{r}^{T}{\bf A}={\bf 1}_{n}^{T}.
\end{align}
Using ADMM, the augmented Lagrangian (AL) can be written as:  
\begin{align}\label{eq: FCLSU3}\nonumber
 & \hat{\bf A},\hat \bS=\arg\min_{{\bf A},{\bS}} \frac{1}{2} || {\bf Y}-{\bf E}{\A}||_{F}^{2}+\frac{\mu}{2} ||\bS-\A-\bL||_{F}^{2}\\ &~~ {\rm s.t.}~~
{\bf S}\geq 0,{\bf 1}_{r}^{T}{\bf A}={\bf 1}_{n}^{T},
\end{align}
where $\bL$ is the Lagrange multiplier. Problem (\ref{eq: FCLSU3}) can be solved w.r.t. ${\bf A} $ and ${\bf S} $, separately. With respect to $\A$ we have a quadratic programming (or least
squares) with the equality constraint as:
\begin{align}\label{eq: QuEC1}\nonumber
 & \hat{\bf A}=\arg\min_{{\bf A}} \frac{1}{2} || {\bf Y}-{\bf E}{\A}||_{F}^{2}+\frac{\mu}{2} ||\bS-\A-\bL||_{F}^{2}\\ &~~ {\rm s.t.}~~{\bf 1}_{r}^{T}{\bf A}={\bf 1}_{n}^{T}
\end{align}
The Lagrangian function is given by:
\begin{equation}\label{eq: QuEC11}
 \Lagr(\A,\nu)=\frac{1}{2} || {\bf Y}-{\bf E}{\A}||_{F}^{2}+\frac{\mu}{2} ||\bS-\A-\bL||_{F}^{2}+\nu^T({\bf 1}_{r}^{T}{\bf A}-{\bf 1}_{n}^{T}).
\end{equation}
Here, we derive the Karush-Kuhn-Tucker (KKT) conditions for (\ref{eq: QuEC11}). For the stationary condition we have 
$\nabla_\A \Lagr(\A,\nu)=0$,
\begin{equation}\label{eq: QuEC22}
 (\E^T\E+\mu\I)\A+\1_r\nu=\E^T\Y+\mu(S-L)
\end{equation}
Therefore, by holding the primal feasibility ${\bf 1}_{r}^{T}{\bf A}={\bf 1}_{n}^{T}$
\begin{equation}
    \begin{pmatrix}
    \E^T\E+\mu\I&\1_r\\
    \1_r^T&0
    \end{pmatrix}
        \begin{pmatrix}
    \A\\
    \nu
    \end{pmatrix}
    =
        \begin{pmatrix}
    \E^T\Y+\mu(\bS-\bL)\\
    \1_n^T
    \end{pmatrix}.
\end{equation}
We use the blockwise inversion to solve for $\A$:
\begin{equation}\label{eq: QuEC33}
 \hat\A=(\Q+\Q\1_r {\textnormal c}\1_r^T\Q)(\E^T\Y+\mu(\bS-\bL)) -\Q\1_r {\textnormal c}\1_n^T,
\end{equation}
where
\begin{align}\label{eq: QuEC344}
 & \Q=(\E^T\E+\mu\I_r)^{-1}\\ &\label{eq: QuEC345} {\textnormal c}=-1/(\1_r^T\Q\1_r).
\end{align}
The augmented term in (\ref{eq: QuEC11}) makes matrix $\Q$ to be always non-singular ($\mu>0$) and hence our solution (\ref{eq: QuEC33}) is feasible. To simplify the algorithm, we demonstrate the solution of the quadratic programming with the equality constraint (QuEC) by:
 \begin{equation}\label{eq: QuEC}
  {\bf A}=QuEC(\A,\bS,\bL;\Y,\E,\mu),
\end{equation}
where $QuEC$ is given by (\ref{eq: QuEC33}), (\ref{eq: QuEC344}), and (\ref{eq: QuEC345}).

With respect to $\A$, the problem turns into: 
\begin{equation}\label{eq: FCLSU6}\nonumber
  \hat \bS=\arg\min_{{\bS}} \frac{\mu}{2} ||\bS-\A-\bL||_{F}^{2}~~ {\rm s.t.}~~
{\bf S}\geq 0,
\end{equation}
and the solution is: 
\begin{equation}
    \label{eq: FUnmix2}
\bS=\max(0,\A+\bL),
\end{equation}
and the multiplier is updated using: 
\begin{equation}
    \label{eq: L}
\bL=\bL+\A-\bS.
\end{equation}


\subsection{${\bf B}$-step:} In the next step, we assume that ${\bf A}$ is fixed, turning problem (\ref{eq: M3}) into:
    \begin{align}\label{eq: Bstep}\nonumber
  &\hat{\bf B}=\arg\min_{{\bf B}} \frac{1}{2} || {\bf Y}-{\bf DBA}||_{F}^{2}+\lambda\left\| {\bf DB} - {\bf m}{\bf 1}_{r}^{T} \right\|_F^2\\ &~~ {\rm s.t.}~~
{\bf B}\geq 0,{\bf 1}_{m}^{T}{\bf B}={\bf 1}_{r}^{T}.
\end{align}
Splitting the variables as $\B=\bS_1$ and $\D\B=\bS_2$: 
    \begin{align}\label{eq: BstepAL}\nonumber
  &\hat{\bf B}=\arg\min_{\B,\bS_1,\bS_2} \frac{1}{2} || {\bf Y}-\bS_2\A||_{F}^{2}++\lambda\left\| {\bf \bS_2} - {\bf m}{\bf 1}_{r}^{T} \right\|_F^2\\ &\nonumber+\frac{\mu}{2}|| \bS_1-\B-\bL_1||_{F}^{2}+\frac{\mu_1}{2}|| \bS_2-\D\B-\bL_2||_{F}^{2}~~ \\ &{\rm s.t.}~~
\bS_2\geq 0,{\bf 1}_{m}^{T}{\bf B}={\bf 1}_{r}^{T}.
\end{align}
We solve (\ref{eq: BstepAL}) with respect to each unknown matrix separately. Therefore, we have: 
\\
\begin{equation}\label{eq: QuEC4}
  \hat\B=QuEC(\bS_1,\bL_1;(\bS_2-\bL_2),\D,\mu_1/\mu_2),
  \end{equation}
\begin{equation}
    \label{eq: BstepS2}
\hat\bS_1=\max(0,\B+\bL_1),
\end{equation}
\begin{equation}
   \hat \bS_2=(\Y\A^T+\lambda{\bf m}{\bf 1}_{r}^{T}+\mu_2(\D\B+\bL_2))(\A\A^T+(\mu_2+\lambda)\I_r)^{-1}.
\end{equation}
Finally, we update the multipliers
\begin{equation}
    \label{eq: L2}
\bL_1=\bL_1+\B-\bS_1,
\end{equation}
\begin{equation}
    \label{eq: L3}
\bL_2=\bL_2+\D\B-\bS_2.
\end{equation}
Here, we initialize $\bS_1$, $\bS_2$, $\bL_1$, and $\bL_2$ with zero. The $\A$-Step and $\B$-Step should be repeated until convergence; otherwise, the cyclic descent with respect to $\A$ and $\B$ may fail due to the non-convex nature of the problem. The pseudo-code for MiSiSUn is provided in Algorithm \ref{Alg:MiSiSUn}. Note that the number of endmembers must be specified for MiSiSUn.

\begin{algorithm}
[tbp]\footnotesize
\SetAlgoLined
\vspace{.cm}
\KwIn{${\bf Y}$: Hyperspectral data, ${\bf D}$: Endmember library, $r$: Number of endmembers, $\mu$, $\mu_1$, and $\mu_2$: AL parameters.}
\vspace{.cm}
\KwOut{$\hat{\bf A}$: Abundances, $\hat{\bf E}$: Endmembers, $\hat{\bf B}$: Endmembers' contributions.}
\textbf{Initialization}: ${\bf B}=0$, ${\bf S}=0$, ${\bf L}=0$, $\bS_i=0$, $\bL_i=0$, $i=1,2$ 

\For{$t = 1$ \KwTo $T$}{
\textbf{A-step} :\\ 
\For{$i=1$ \KwTo $T_1$}{
 $\A=QuEC(\A,\bS,\bL;\Y,\D\B,\mu)$
 \\
$\bS=\max(0,\A+\bL)$\\
$\bL=\bL+\A-\bS$
}
\textbf{B-step} :\\ 
\For{$i=1$ \KwTo $T_2$}{$\B=QuEC(\bS_1,\bL_1;(\bS_2-\bL_2),\D,\mu_1/\mu_2)$\\
$\bS_1=\max(0,\B+\bL_1)$\\
$\bS_2=(\Y\A^T+\lambda{\bf m}{\bf 1}_{r}^{T}+\mu_2(\D\B+\bL_2))(\A\A^T+(\mu_2+\lambda)\I_r)^{-1}$\\
$\bL_1=\bL_1+\B-\bS_1$\\
$\bL_2=\bL_2+\D\B-\bS_2$
}
}
$\hat{\E} = \D \hat{\B}$
\caption{MiSiSUn}
\label{Alg:MiSiSUn}
\end{algorithm}

\section{Experimental Results}

We evaluated and compared the performance of the proposed method with eight semi-supervised unmixing methods. Among them, four methods belong to the SRM category (SUnSAL, CLSUnSAL, MUA\_SLIC, and S2WSU), one is a deep-learning-based method (SUnCNN), and the remaining three belong to the AAM category (SUnAA, SUnS, and FaSUn). 
The source code used for running SUnSAL, CLSUnSAL, MUA\_SLIC, S2WSU, and SUnCNN is available at  \href{https://github.com/BehnoodRasti/HySUPP}{https://github.com/BehnoodRasti/HySUPP}, and  SUnS and FaSUn were provided in a dedicated Python package called Fast Semisupervised Unmixing (FUnmix) available at \href{https://github.com/BehnoodRasti/FUnmix}{https://github.com/BehnoodRasti/FUnmix}. For the sake of reproducibility, we refactored the code and distributed it as a standalone library-based unmixing Python package, available at \href{https://github.com/BehnoodRasti/MiSiSUn}{https://github.com/BehnoodRasti/MiSiSUn}. 
The hyperparameters for the selected methods were fine-tuned and shown in Table \ref{tab:hparams}.

For a quantitative evaluation of the performance of the abundance estimation, we employed the signal-reconstruction-error (SRE) measured in decibels (dB), defined by:
\begin{equation} \label{eq:SRE}
    \text{SRE}(\A, \hat{\A})=20\log_{10}\frac{\|{\A}\|_F}{\|{\A}-\hat{\A}\|_F}.
\end{equation}

\begin{table*}[h]
    \centering
        \caption{Hyperparameters used in the experiments in different scenarios.}
    \begin{tabular}{c|cc}
    \toprule
     Methods    & Simulated & Cuprite\\
    \midrule
      SUnSAL   & SNR-dependent (see \href{https://github.com/ricardoborsoi/MUA\_SparseUnmixing}{source code}) & $\lambda = 0.005$\\
      CLSUnSAL   & SNR-dependent (see \href{https://github.com/ricardoborsoi/MUA\_SparseUnmixing}{source code}) & $\lambda = 0.05, \mu=0.01$\\
      MUA\_SLIC   & SNR-dependent (see \href{https://github.com/ricardoborsoi/MUA\_SparseUnmixing}{source code})  & $\lambda_1 = 0.001, \lambda_2 = 0.01, \beta=10, \text{slic\_size}=200, \text{slic\_reg}=0.01$\\
      S2WSU   & SNR-dependent (see \href{https://github.com/ricardoborsoi/MUA\_SparseUnmixing}{source code}) & $\lambda = 0.001$\\
      SUnCNN   & SNR-dependent (see \href{https://github.com/BehnoodRasti/SUnCNN}{source code} & $\text{niters}=20000$\\
      SUnAA   & default (see \href{https://github.com/inria-thoth/SUnAA}{source code})& default \\
      SUnS   & $T=10000, T_A = T_B = 5, \mu_1 = 50, \mu_2 = 2, \mu_3 = 1, \lambda=0.01$ & $T=10000, T_A = T_B = 5, \mu_1 = 400, \mu_2 = 100, \mu_3 = 1, \lambda=0.1$\\
      FaSUn   & $T=10000, T_A = T_B = 5, \mu_1 = 50, \mu_2 = 2, \mu_3 = 1$ & $T=10000, T_A = T_B = 5, \mu_1 = 400, \mu_2 = 20, \mu_3 = 1$ \\
      MiSiSUn   & $T=10000, T_A = T_B = 5, \mu_1 = 50, \mu_2 = 2, \mu_3 = 1, \lambda=0.3$ & $T=10000, T_A = T_B = 5, \mu_1 = 500, \mu_2 = 50, \mu_3 = 1, \lambda=10$ \\
    \bottomrule
    \end{tabular}
    \label{tab:hparams}
\end{table*}

\subsection{Data Description}
We have performed three different types of experiments that require different datasets. The first dataset validates the effectiveness of the method in different noisy scenarios. The second dataset evaluates the potential of the method when pixel purity varies, while the third dataset confirms its effectiveness in real-world scenarios.


\subsubsection{Sim1 dataset: Synthetic Datasets with Spatial Structure}
\begin{figure}[htbp]
\centering
  \begin{tabular}{cc}
    \includegraphics[width=.21\textwidth]{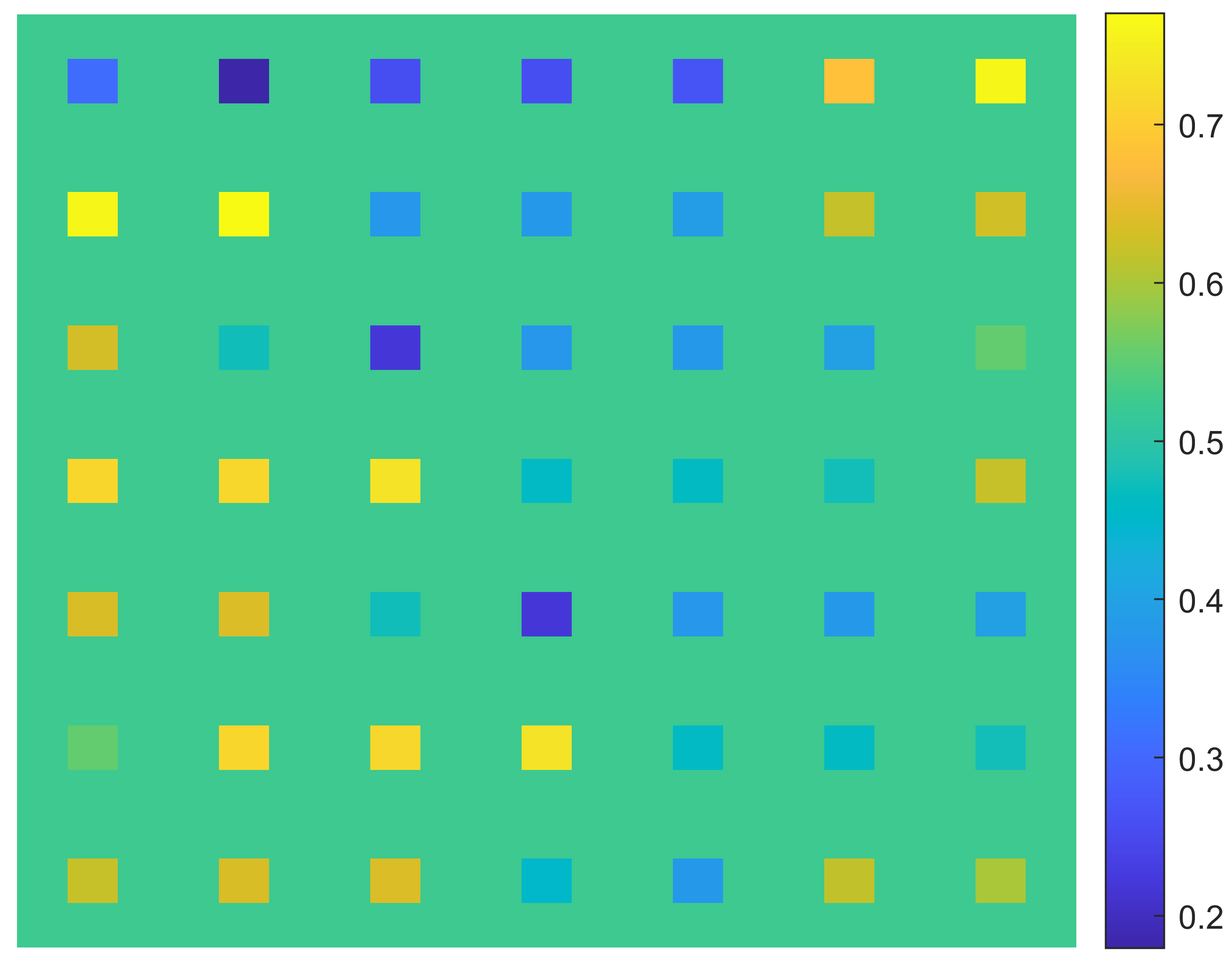}& \includegraphics[width=.24\textwidth]{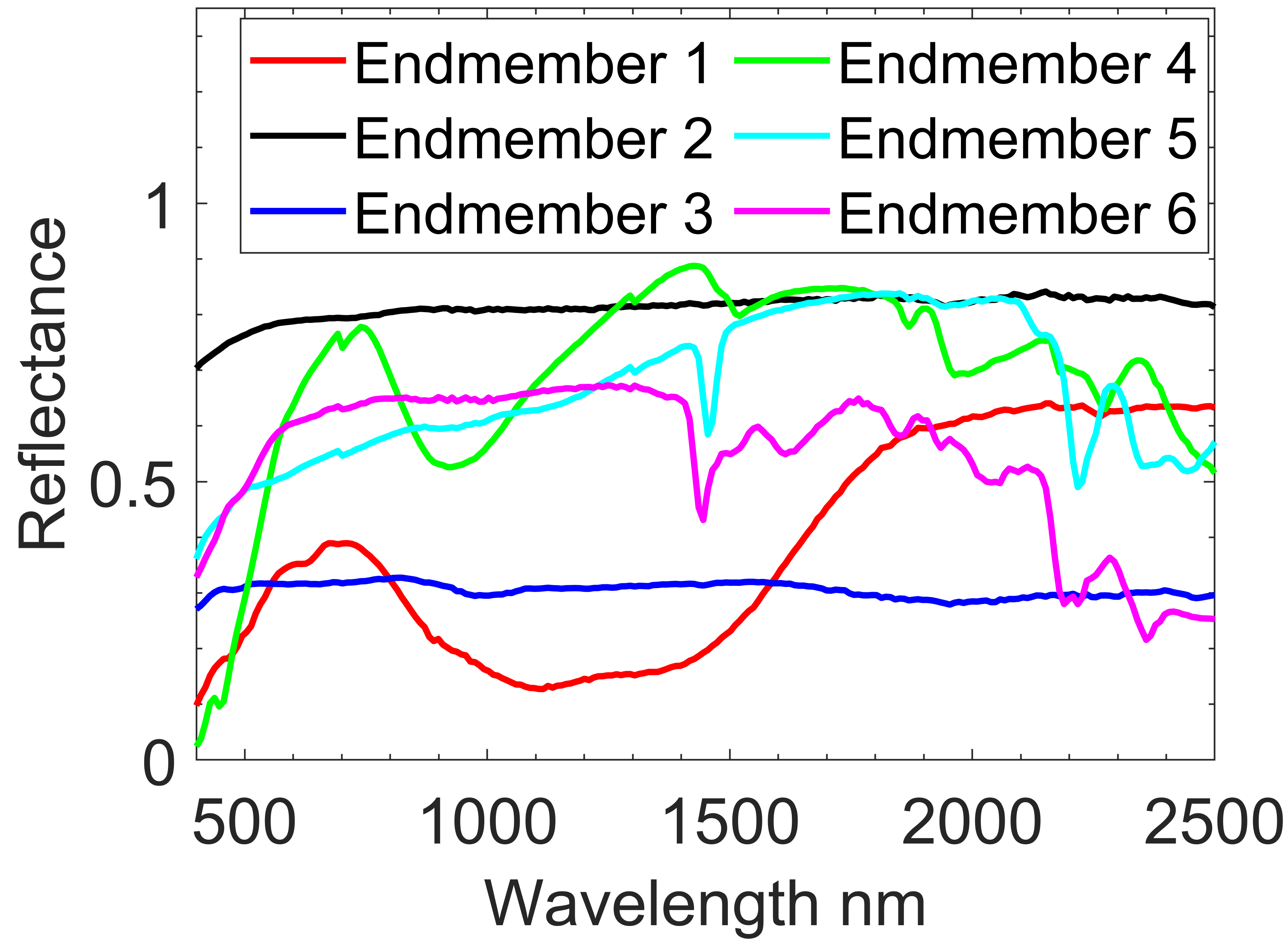}\\ 
    (a) & (b)\\
  \end{tabular}
  \caption{Sim1 dataset: a) Band number 70 {\color{black} (1050 nm)}  b) Endmembers.}
\label{Simulated Dataset 1}
\end{figure}

A simulated hyperspectral dataset comprising 105 $\times$ 105 pixels (see Fig. \ref{Simulated Dataset 1}(a)) is generated through a linear combination of six endmembers. The endmembers of this dataset are shown in Fig. \ref{Simulated Dataset 1}(b) and contain 224 reflectance values within the wavelength range of [400–2500] nm. This hyperspectral image contains 49 homogeneous squares, each measuring $5\times5$ pixels. Of these, 45 squares represent binary mixtures (for each endmember pair, three binary mixtures (0.75/0.25, 0.5/0.5, and 0.5/0.5) were considered), while the remaining four correspond to randomly generated higher-order mixtures. All background pixels share identical fractional abundances, given by $1/6 \times [1,1,1,1,1,1]^T$.  We clarify that pure pixels are not present in the dataset, and the maximum abundance per material is limited to 75\%.

To demonstrate the proposed method's potential in noisy scenarios, Gaussian noise is added to achieve different signal-to-noise ratios (SNRs) (i.e., 20, 30, and 40 dB).

\subsubsection{Sim2 dataset: Synthetic Datasets with varying Pixel Purity Levels}
\begin{figure}[htbp!]
\centering
  \begin{tabular}{cc}
    \includegraphics[width=.40\textwidth]{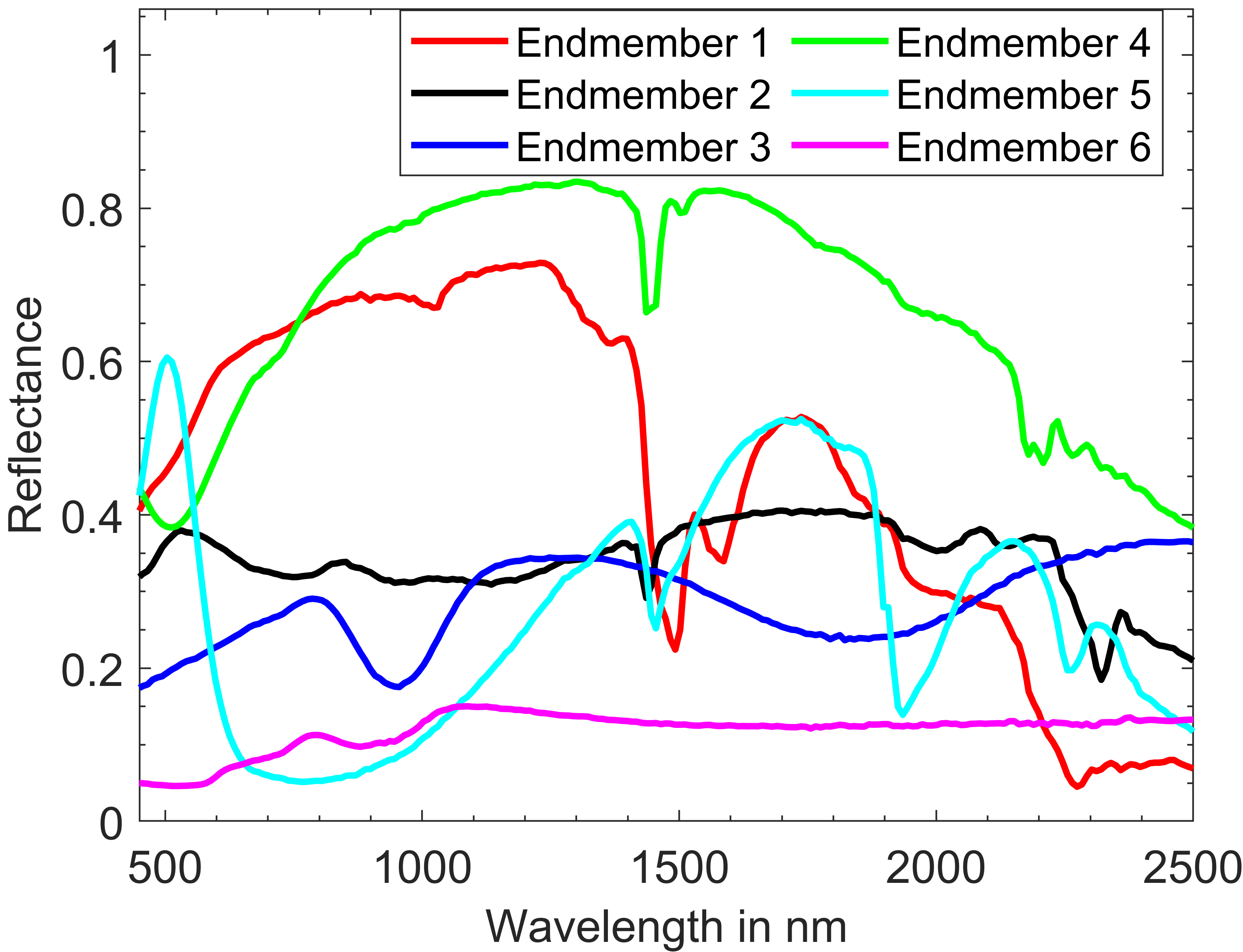}\\ 
  \end{tabular}
  \caption{Endmembers used to simulate the Sim2 dataset.}
\label{Simulated Dataset 2}
\end{figure}

Similar to the Sim1 dataset, six spectra were selected from the USGS library (see Fig. \ref{Simulated Dataset 2}) and linearly combined to create a dataset of size $100 \times 100$ pixels. This dataset does not contain a spatial structure but instead is parameterized by a pixel purity level. The degree of pixel purity was quantified by the parameter $\rho$, with lower values signifying lower purity and higher values indicating larger purity. The fractional abundances used to generate this dataset were initially sampled from a symmetric Dirichlet distribution, and those with values exceeding $\rho$ were filtered out. 
Synthetic Gaussian noise is eventually added to obtain an input SNR of 30 dB.

\subsubsection{Cuprite Dataset}
The Cuprite dataset used in this paper contains 250 $\times$ 191 pixels. Similar to the Sim1 and Sim2 datasets, each spectrum contains 224 reflectance values within the wavelength range of [400–2500] nm. Note that we remove the water absorption and noisy bands ([1-2,104-113,148-167,221-224]), such that the final pixels are of dimension $p=188$. Cuprite is a well-studied mineral site, for which geological ground reference is available for the dominant minerals (see Fig. \ref{fig:Cuprite} (a)). In this scene, 12 materials are available, of which  Alunite, Chalcedony, Kaolinite, Montmorillonite, and Sphene are the most dominant. The GT abundance maps of these minerals (see the first column of \ref{fig:Cuprite} (b)) were produced through supervised classification using the geological reference map of Fig. \ref{fig:Cuprite} (a). For this purpose, a number of training samples were manually selected, and the random forest classifier was applied to train the model and to classify all remaining test pixels.


For all experiments on both the simulated and real datasets, and for all the comparing methods, we use the same library composed of 498 spectra obtained from the USGS library for semi-supervised unmixing.

\subsection{Experimental Results: Synthetic Datasets}

For the simulated datasets, we performed five independent runs and then averaged the results. Standard deviations are indicated by error bars.

\begin{figure*}[htb!]
  \centering
  {\includegraphics[width=.95\textwidth]{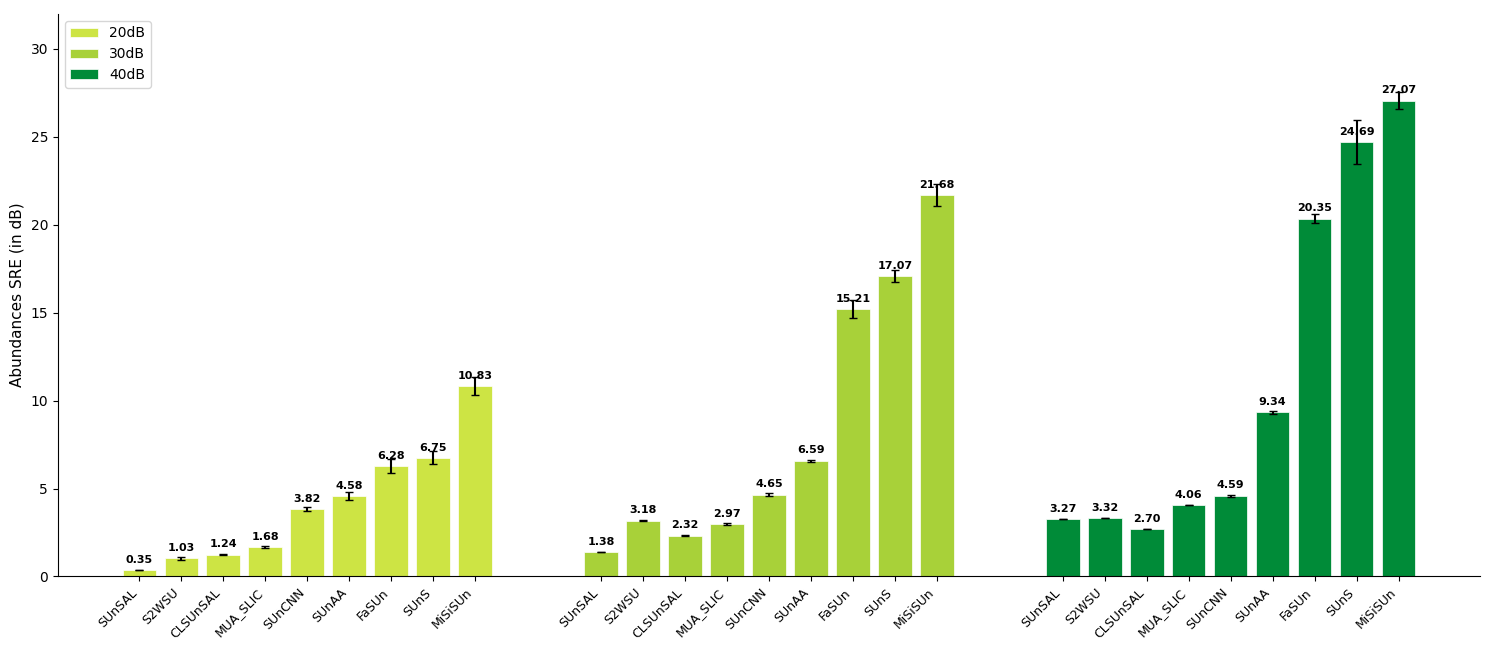}}
  \caption{Abundance SRE in dB obtained by the selected state-of-the-art methods on the Sim1 dataset for different noise levels.}
  \label{fig:sim1}
\end{figure*}

In the first experiment, we compare the performance of the proposed method with the state-of-the-art approaches on the Sim1 dataset under different levels of noise. Fig. \ref{fig:sim1} summarizes the results in terms of SRE for 20, 30, and 40 dB of noise. The following observations can be made:
\begin{itemize}
\item All methods exhibit improved performance as the SNR increases.
\item Low-rank methods (i.e., SUnAA, FaSUn, and SUnS) significantly outperform the methods SUnSAL, S2WSU,  MUA\_SLIC, SUnCNN, and CLSUnSAL.
\item The proposed method (MiSiSUn) achieves the best performance across all noise levels. This is due to the efficient integration of the properties of AAM and the volume constraint enforced on the linear simplex. The performance gap between the proposed method and the competing approaches becomes more pronounced at lower SNR levels.
\end{itemize}

\begin{figure*}[htb!]
  \centering
{\includegraphics[width=.95\textwidth]{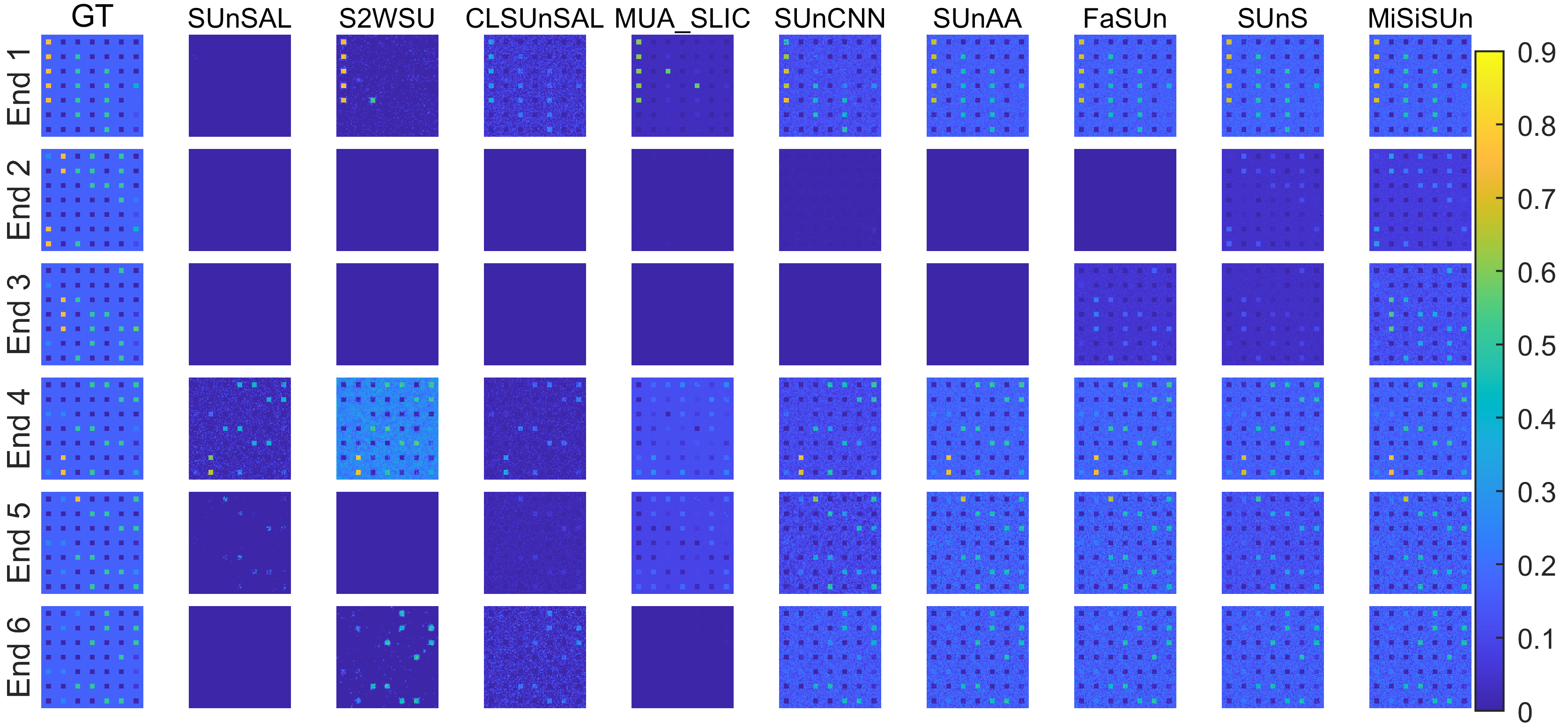}}
  \caption{Estimated abundances by the selected state-of-the-art methods on the  Sim1 dataset for a given input SNR (20 dB).}
  \label{fig:Sim1_Abundance}
\end{figure*}

Fig. \ref{fig:Sim1_Abundance} presents a visual comparison of the estimated abundance maps on the Sim1 dataset at an SNR of 20 dB. Overall, the results indicate that MiSiSUn achieves the best performance, particularly for endmembers 2 and 3, where most other methods fail. On the other endmembers, FaSUn and SUnS exhibit comparable performance. while the abundance maps produced by SUnCNN and SUnAA appear visually similar. In contrast, MUA\_SLIC generates oversmoothed abundance maps, likely due to the prior segmentation step.

\begin{figure*}[h]
  \centering
  {\includegraphics[width=.95\textwidth]{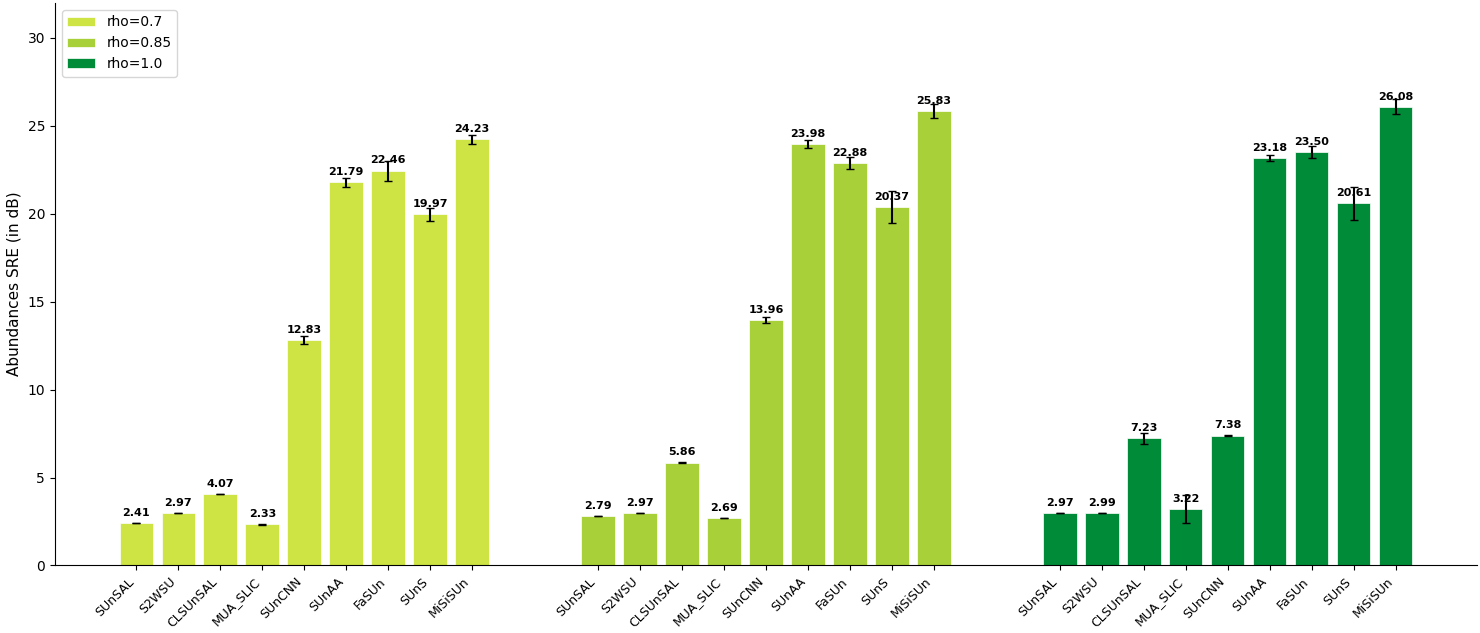}}
  \caption{Abundance SRE in dB obtained by the selected state-of-the-art methods on the Sim2 dataset at a given input SNR (30 dB) for different pixel purity levels.}
  \label{fig:sim2}
\end{figure*}

Fig. \ref{fig:sim2} summarizes the results obtained by the state-of-the-art methods on the Sim2 dataset. As observed, the performance of most approaches improves as pixel purity increases. Consistent with the results on the Sim1 dataset, MiSiSUn outperforms the competing methods on this dataset, demonstrating the superior performance of the proposed approach in highly mixed scenarios. 
SunAA and FaSUn achieve similar performance and are slightly superior to that of SUnS. In contrast, SUnSAL performs significantly worse, with results at least 8.5 times lower than those of MiSiSUn. Interestingly, SUnCNN performs better when the data is highly mixed compared to the scenarios with purer pixels.

\subsection{Experimental Results: Real Data}

\begin{figure*}[htb!]
\centering
   \subfloat[Reference Map]{\includegraphics[width=.16\textwidth]{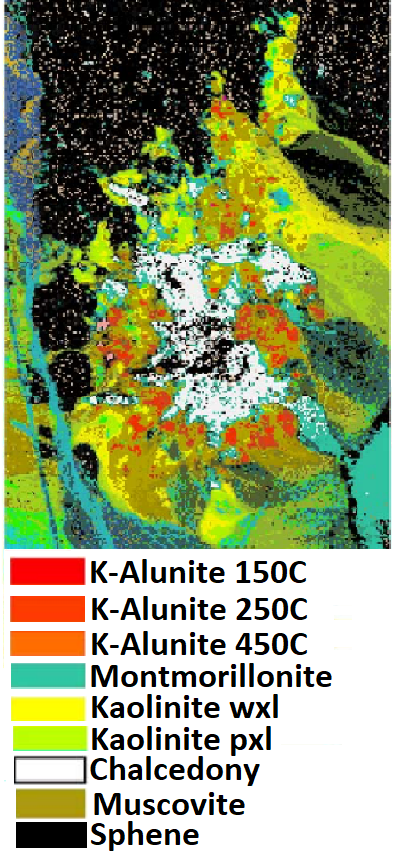}}
   \subfloat[Estimated abundances]{\includegraphics[width=0.8\textwidth]{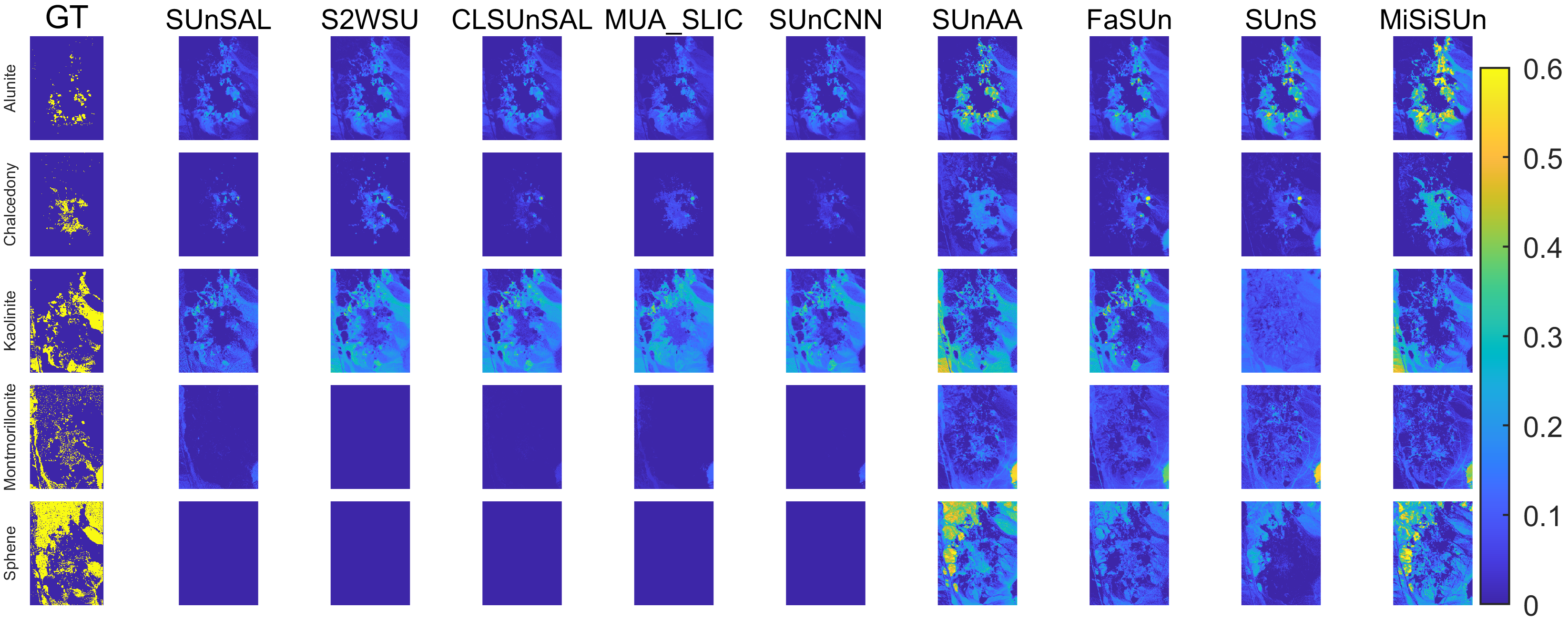}}
  \caption{Abundance maps obtained by the different methods on Cuprite,  along with the geological reference map and the produced GT abundance maps from this map.}
\label{fig:Cuprite}
\end{figure*}

Fig. \ref{fig:Cuprite}(b) shows the estimated abundance maps for five dominant materials, i.e.,  Alunite, Chalcedony, Kaolinite, Montmorillonite, and Sphene, from the Cuprite dataset obtained by the different state-of-the-art methods. The number of endmembers in the scene (i.e., $r$) was set to 12. As no ground-truth fractional abundances are available for this dataset, the geological map and the obtained GT abundance maps from this map are used as reference (see Figs. \ref{fig:Cuprite}(a) and (b)) for visual comparison.

As expected, the low-rank methods SUnAA, FaSUn, SUnS, and MiSiSUn estimated meaningful abundance maps. Among the low-rank methods, MiSiSUn performs best, especially for Alunite and Chalcedony. Abundance maps produced by the conventional sparse methods contain very low abundance values due to the distribution of abundances over multiple endmembers. 

\begin{figure*}[htb!]
\centering
{\includegraphics[width=0.8\textwidth]{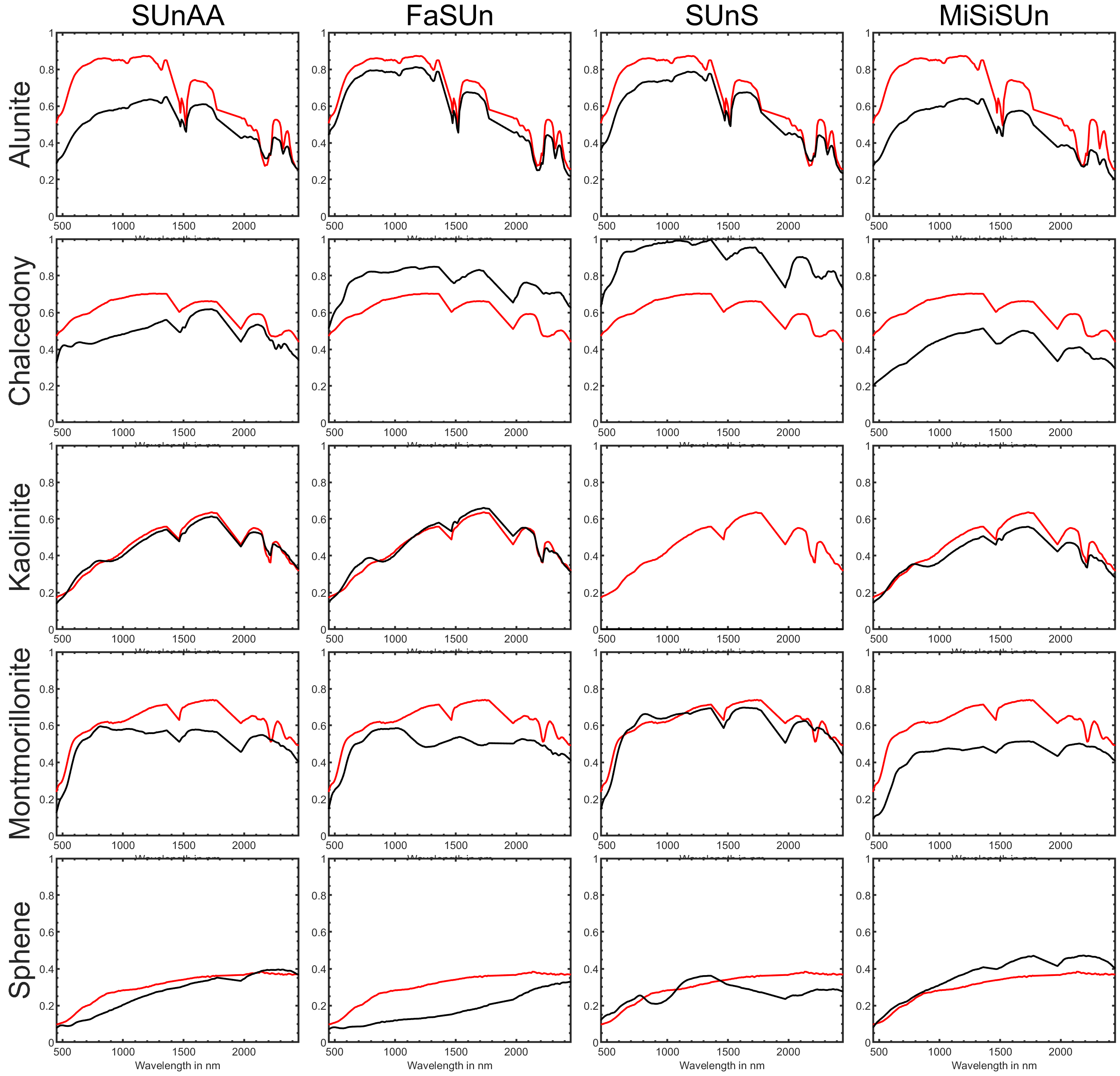}}
  \caption{Endmembers obtained by the different low-rank methods to Cuprite, compared with the USGS spectral library. The ground-truth endmembers are depicted in red, the estimated ones in black.}
\label{fig:Cuprite_endmembers}
\end{figure*}

Fig. \ref{fig:Cuprite_endmembers} compares the library endmembers (red solid lines) with the estimated endmembers (black solid lines) for the five dominant materials in the Cuprite dataset: Alunite, Chalcedony, Kaolinite, Montmorillonite, and Sphene. Since the spectral library contains multiple spectra per material, the library endmembers shown were selected based on their ability to produce abundance maps (using fully constrained least squares unmixing) that are visually consistent with the geological reference map. Table \ref{tab:SAD} reports the Spectral Angle Distances (SAD). The results demonstrate that, among the low-rank methods, SUnS performs the worst for Kaolinite, with a zero vector for the estimated endmember, whereas SUnAA estimates it perfectly. SUnS accurately estimates the  Montmorillonite endmember, while the endmember estimated by FaSUn deviates significantly from the spectral shape of the ground truth. Although SUnS accurately estimates the spectral shape of Chalcedony, the estimated abundance map is less impressive than that estimated by MiSiSUn. All methods fail to predict the spectral shape of the featureless mineral Sphene; however, the endmember estimated by MiSiSUn more closely resembles the reference map. Although FaSUn accurately estimates the spectrum of Alunite, the corresponding abundance map is not as satisfactory as the ones obtained by SUnAA and MiSiSUn, when compared with the reference map. This discrepancy could be attributed to a mismatch between the library endmember and the measured hyperspectral dataset.

\begin{table}[]
\caption{SAD between estimated and library endmembers for the Cuprite dataset. The Best Performances are Shown in Bold.}
\begin{tabular}{lllll}
\hline
SAD (in degree) & SUnAA & FaSUn & SUns  & MiSiSUn \\
\hline
Alunite         & 7.68  & \textbf{1.48}  & 3.68  & 5.97    \\
\hline
Chalcedony      & 6.09  & 3.43  & \textbf{3.06}  & 7.33    \\
\hline
Kaolinite       & \textbf{2.46}  & 2.67  & 16.59 & 2.78    \\
\hline
Montmorillonite & 4.83  & 6.55  & \textbf{3.91}  & 6.53    \\
\hline
Sphene          & 8.26  & 16.23 & 10.92 & \textbf{3.45}   \\
\hline
\end{tabular}
\label{tab:SAD}
\end{table}

\subsection{Processing time}

Table \ref{tab:timer} compares the processing time (in seconds) of the unmixing algorithms applied to the different datasets. For consistency, all reported processing times are mean values over five experiments. Sim1\_extend is the extended version of Sim1 with an increased number of pixels (around nine times more pixels) to show the scalability of the models. All methods were implemented in Python (3.10.11), and the results were obtained on a computer with an Intel Core i9-12900KF processor (3.19 GHz), 64 GB of memory, a 64-bit Operating System, and an NVIDIA GeForce RTX 3080 graphics processing unit.  

As can be seen, processing time increases with the number of pixels for all methods. MUA\_SLIC is the most efficient method for the Sim1 dataset, while SUnSAL requires the least processing time for the Sim2 and Cuprite datasets. The processing times of SUnS, FaSUn, and MiSiSUn do not differ significantly, which confirms the scalability of those algorithms. SUnAA is slightly slower than the competing low-rank methods, due to the columnwise update of ${\bf B}$. SUnCNN is the slowest method for the Cuprite dataset, due to the high number of iterations suggested for this dataset. We should note that among the methods mentioned above, SUnCNN, FaSUn, SUnS, and MiSiSUn
use a GPU, while the others use a CPU for their computations. These results make clear that sparse unmixing methods remain difficult to scale. It is worth mentioning that the ADMM update in sparse unmixing models is $\mathcal{O}\big(nm^2\big)$ (excluding the one-time computation of ${\bf D}^T{\bf D}$) and typically requires solving a linear system involving an $m\times m$ matrix and therefore the size of the library limits the scalability of sparse unmixing algorithms. On the other hand, the computational complexity for each iteration of the MiSiSUn, FaSUn, and SUnS algorithms is $\mathcal{O}\big((T_1+T_2)\times  n\times p \times 
r\big)$, which makes them capable of working with a large library.
\begin{table*}[htb!]
\centering
\caption{Processing time. The best results are in boldface, and the second best are underlined.}
\resizebox{\textwidth}{!}{
\begin{tabular}{c|c|c|c|c|ccccccccc}
\toprule
             & \# Pixels & \# Bands & \# Endmembers & \# Atoms & SUnSAL & S2WSU & CLSUnSAL & MUA\_SLIC& SUnCNN & SUnAA  & FaSUn & SUnS & MiSiSUn\\
\midrule
Sim1          &      11025     &   224       &      6        &    498    &    \underline{41.6}      &     107.4      &      55.8 & \textbf{37.1}   &   68.0      & 189.5 & 126.4  & 129.5  & 122.0  \\
Sim2          &      10000     &    224      &      6         &    498    &    \textbf{20.5}      &      84.7     &    56.9 & \underline{38.9}     &   60.0    &      159.0  &  111.3  & 115.3   &  113.2\\
Cuprite      &     47750      &     188     &      12         &    498    &   \textbf{47.7}       &    321.4       &    175.6 & \underline{116.9}    &   565.8    &      {195.5} & {134.4} & 134.0   &  138.7   \\
Sim1\_extend      &     99225      &     224     &      6         &    498    &   {339.6}       &    884.2       &    341.8 & {325.2}      &   477.4    &    {1631.4} & \textbf{148.9} &  176.6   &  \underline{152.5}   \\
\bottomrule
\end{tabular}
}
\label{tab:timer}
\end{table*}

\section{Conclusion}
We proposed a novel library-based (semisupervised) unmixing method for hyperspectral data, called MiSiSUn. Unlike the sparse unmixing models, MiSiSUn incorporates the properties of AAM and the linear simplex to accurately estimate the fractional abundances of the highly mixed datasets. Additionally, the proposed ADMM-based algorithm shows promising scalability features. The results on two simulated datasets and Cuprite confirm the advantages of the proposed method compared to the state of the art, both in terms of abundance estimation and efficiency. Moreover, we showed that the estimated endmembers for the Cuprite dataset match the measured ones from the library.

\section*{Acknowledgment}
The research presented in this paper is funded by the Research Foundation-Flanders (project G031921N). The work of Bikram Koirala was supported by the Research Foundation Flanders, Belgium (FWO: 1250824N-7028). The authors thank Dr. Alexandre Zouaoui for providing the codes for the competitive methods.  

\ifCLASSOPTIONcaptionsoff
  \newpage
\fi



%
\bibliographystyle{IEEEbib}
\bibliography{New, refs,new_ref,refs1,HyDe,refs_misicnet,Ref_non}

%








\end{document}